# UV soliton dynamics and Raman-enhanced super-continuum generation in photonic crystal fiber

Pooria Hosseini,* Alexey Ermolov, Francesco Tani, David Novoa, and Philip St.J. Russell

*Max Planck Institute for the Science of Light, Staudtstrasse 2, 91058 Erlangen, Germany*
*Corresponding author: pooria.hosseini@mpl.mpg.de*

Ultrafast broadband ultraviolet radiation is of importance in spectroscopy and photochemistry, since high photon energies enable single-photon excitations and ultrashort pulses allow time-resolved studies. Here we report the use of gas-filled hollow-core photonic crystal fibers (HC-PCFs) for efficient ultrafast nonlinear optics in the ultraviolet. Soliton self-compression of 400 nm pulses of (unprecedentedly low) ~500 nJ energies down to sub-6-fs durations is achieved, as well as resonant emission of tunable dispersive waves from these solitons. In addition, we discuss the generation of a flat supercontinuum extending from the deep ultraviolet to the visible in a hydrogen-filled HC-PCF. Comparisons with argon-filled fibers show that the enhanced Raman gain at high frequencies makes the hydrogen system more efficient. As HC-PCF technology develops, we expect these fiber-based ultraviolet sources to lead to new applications.

**Keywords:** *ultraviolet, pulse compression, soliton, super continuum, photonic crystal fiber, Raman scattering*

Deep and vacuum ultraviolet (DUV-VUV) sources are highly relevant in spectroscopy and photochemistry, since the majority of molecules have absorption bands in this spectral range.[1,2] Although, broadband DUV light has been generated in bulk materials such as $CaF_2$[3,4] and optical fibers with soft-glass cores,[5] the VUV is much more difficult to access because of the limited transparency, strong dispersion and damage thresholds of most materials.

Ultrafast sources in the UV permit time-resolved capture of the electronic and vibrational dynamics of molecules.[6-8] Several approaches to generating ultrashort UV pulses exist, such as harmonic generation,[9] sum-frequency mixing,[10] achromatic frequency doubling[11] and spectral broadening in gas-filled capillaries.[12] Most of these techniques rely, however, on components such as gratings, prisms, dispersive and deformable mirrors to compensate the spectral phase, so suffer from optical loss and present limitations in the few-cycle UV regime.

By side-stepping these limitations, gas-filled broadband-guiding hollow-core photonic crystal fiber (HC-PCF) provides an ideal platform for UV nonlinear optics, combining broad UV transparency and pressure-tunable shallow dispersion with a very high damage threshold.[13] Further, by tightly confining the light in a small core, the pulse energies needed for phenomena such as soliton self-compression[14] and supercontinuum generation[15] are reduced to the few µJ range, compared to the mJ energies needed in conventional set-ups that employ wide-bore capillaries. This uniquely permits repetition rate scaling up to the multi-MHz range at pulse energies compatible with current 1 µm fiber lasers.[16]

As a proof of concept, we report the first soliton-based compression of 400 nm pulses, of 40 fs, ~500 nJ down to sub-6-fs durations using an Ar-filled kagomé-type HC-PCF. In contrast to the approaches mentioned above, these few-cycle UV pulses emerge nearly transform-limited from the fiber end-face. This technique is expected to be transferable to the DUV using HC-PCFs with smaller core-sizes and thinner core-walls (to avoid loss-inducing anti-crossings between the core mode and resonances in the core walls).

We provide further evidence of the excellent UV performance of these fibers[17] in the observation of dispersive wave (DW) emission by self-compressed UV solitons. These DWs are tunable from the DUV to the VUV by adjusting the filling gas pressure.[18] We were able to generate a DUV-to-visible supercontinuum in a hydrogen-filled HC-PCF pumped at 400 nm. Interestingly, the spectrum is flat to 8 dB from 280 nm to 700 nm, whereas it is much less flat when near-infrared pump pulses are used.[19] We also find that this arrangement outperforms Ar-filled HC-PCF, due to enhanced Raman response in the UV spectral region.

## THEORETICAL BACKGROUND

In fiber-based systems with anomalous dispersion and a focusing Kerr nonlinearity, soliton self-compression occurs for pulses of soliton order $N = (L_D / L_{NL})^{0.5} = (\gamma P_0 T_0^2 / |\beta_2|)^{0.5}$ greater than 1, where $L_D = T_0^2 |\beta_2|^{-1}$ is the dispersion length, $L_{NL} = (\gamma P_0)^{-1}$ the nonlinear length, $\gamma$ the nonlinear fiber parameter, $P_0$ the peak power, $T_0$ the pump pulse duration and $\beta_2$ the group velocity dispersion (GVD).[20] Under these conditions, new frequencies generated by self-phase modulation (SPM) are brought together in time by the negative GVD, causing the pulse to self-compress by a factor $F \sim 4.6N$ over a propagation distance $L_C = L_D (\sqrt{2}N)^{-1}$.[21,22] Soliton dynamics in gas-filled HC-PCF has previously been used to compress ~30 fs infrared pulses with energies of a few µJ down to a single cycle.[23-25]

The $\beta_2$ coefficient of the fiber can be derived from the refractive index of the $LP_{ij}$ mode of broadband-guiding HC-PCFs, which can be accurately estimated using a capillary model:

$$n_{ij}(\omega, p) = \sqrt{n_g^2(\omega, p) - c^2 u_{ij}^2 \omega^{-2} a^{-2}(\omega)}, \qquad (1)$$

where $\omega$ is the angular frequency of the light, $n_g$ the refractive index of the filling gas, $p$ the gas pressure, $u_{ij}$ the $j$-th zero of the $J_i$ Bessel function and $a$ the frequency-dependent effective core radius.[26] The anomalous dispersion of the evacuated hollow core is counterbalanced by the pressure-adjustable normal dispersion of the gas, allowing access to soliton dynamics from the UV to the infrared. At UV wavelengths the gas dispersion is also stronger, forcing the core size to be scaled down if anomalous dispersion is to be maintained.[21]

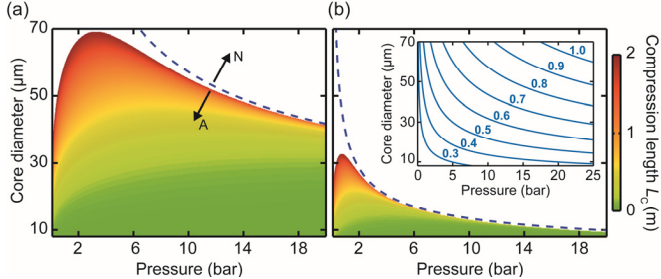

Fig. 1. Analytical comparison of the compression length required for a fixed compression factor of 23 ($N = 5$) for (a) 800 nm and (b) 400 nm pump wavelength. $T_0, P_0$ are kept constant. In the inset the GVD is anomalous below the blue curves, which mark the zero dispersion points at fixed pump wavelength (indicated in µm).

In Fig. 1 we plot the compression length as a function of core diameter and gas pressure at constant soliton order ($N = 5$) and compression factor ($F = 23$), keeping both pulse duration and input power constant, in an Ar-filled HC-PCF pumped at 800 nm (Fig. 1(a)) and 400 nm (Fig. 1(b)). The range of parameters where the dispersion is anomalous is much more restricted for 400 nm pumping. This means that, as anticipated above, the core size has to be kept small if the compression length is to be kept below 1 m. Broadband UV-guiding HC-PCFs with smaller core diameters and low loss (something technically challenging in itself[17,27,28]) will therefore be required for successful use of soliton dynamics in the UV. Additionally, anti-crossings between core modes and core-wall resonances cause narrow bands of high loss which can impair the compression quality by altering the dispersion. They can be pushed spectrally far away from the pump frequency by making the core walls sufficiently thin.[29]

## EXPERIMENTAL SET-UP

A sketch of the experimental set-up is given in Fig. 2(a). A fraction of the energy (120 µJ) from an amplified Ti:sapphire laser system delivering 40 fs pulses at 805 nm was frequency doubled in a 100 µm thick beta-barium-borate (BBO) crystal. The second harmonic at 402 nm (~40 fs duration and up to 10 µJ energy) was separated from the near-infrared parent pulse by a pair of dichroic mirrors, after which a spatial filter was used both to clean the beam profile and to stabilize the beam pointing. Two pairs of chirped mirrors, providing a total group delay dispersion of –800 fs$^2$ (–50 fs$^2$ per bounce at 400 nm), were used to compensate for the dispersion introduced by the optical elements and the air path. A combination of thin-film polarizer and half-wave plate was used to control the power. All the lenses and gas-cell windows were made from uncoated MgF$_2$ glass, which is transparent down to 130 nm and has very low GVD in the UV-visible. The 805 nm pulses were characterized using a commercial GRENOUILLE (Fig. 2(b)), while the 400 nm pulses were measured using a home-built device based on self-diffraction frequency-resolved optical gating (SD-FROG) (Fig. 2(c)), using a 170-µm-thick fused-silica sample as nonlinear medium.[30] The 400 nm pulses were then launched with 70% coupling efficiency into a 70-cm-long kagomé-type HC-PCF with a flat-to-flat core diameter of 22 µm, placed between two gas cells. Figure 2(d) shows a scanning electron micrograph of the fiber microstructure. The core-wall thickness is ~93 nm (Fig. 2(d)), resulting in an anti-crossing[31] at 215-220 nm, far enough from the pump wavelength to allow clean compression.[29] An aluminum off-axis parabolic mirror was used to collimate the fiber output, which was spectrally characterized using a calibrated spectrometer and temporally by the SD-FROG (Fig. 2(c)).

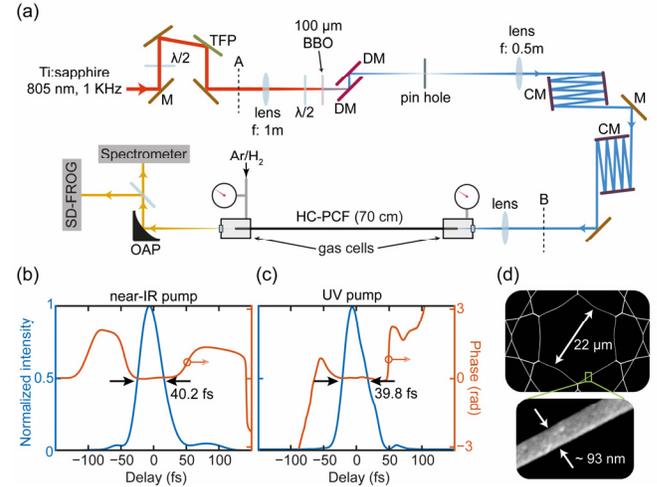

Fig. 2. (a) Experimental set-up. CM, chirped mirror; TFP, thin-film polarizer; DM, dichroic mirror; OAP, off-axis parabolic mirror; M, mirror (dielectric). (b-c) Temporal pulse shape and phase of the (b) near-IR (infrared) beam measured at the position marked with "A" in (a) and (c) UV beam measured at the position marked with "B" in (a). The fiber is evacuated before being filled with Ar or H$_2$. (d) Scanning electron micrograph of the fiber structure. The inset shows the thickness of the fiber core-wall.

## EXPERIMENTAL RESULTS AND DISCUSSION

### A. Generation of sub-6-fs UV pulses at 400 nm

Figure 3(a) shows an SD-FROG trace of a compressed pulse measured at the output of the fiber when filled with 0.8 bar argon and pumped with 520 nJ pulses (measured at the output of the fiber). These parameters give $N \sim 5.5$, providing a good balance between compression factor and quality, while theoretically yielding single-cycle pulses (1.3 fs for a 400 nm pulse). Figure 3(b) shows the reconstructed spectrogram, which is in good agreement with the measurement. On its way to being measured, the self-compressed UV pulse broadens in time (Fig. 3(c)) as a result of the strong chirp it acquires upon transmission through the output window of the gas cell, the air path and the optical elements. The pulse duration at the fiber output face (Fig. 3(d)) was retrieved by numerically subtracting the acquired spectral phase, yielding ~5.7 fs. Note that the main peak carries ~50% of the total pulse energy. The broader reference pulse spectrum, measured before entering the SD-FROG (shown by the dotted curve in Fig. 3(e)), suggests that the Fourier transform-limit of the pulse should be ~4 fs. The discrepancy originates from phase-matching restrictions at short wavelengths in the SD-FROG system. A thinner nonlinear sample and a shallower input angle would provide a larger phase-matching bandwidth, but at the expense of a lower signal-to-noise ratio. Therefore, a compromise between sensitivity and bandwidth has to be made.[30] Compression ratios greater than 7 were reached, comparable to other reports in different spectral regions[23-25] and ultimately limited by third order nonlinear spectral phase in the 805 nm pump pulses (see Figs. 2(b)-(c)). Note that shorter transform-limited pump pulses are likely to improve the quality of the compression, resulting in single-cycle UV pulses.

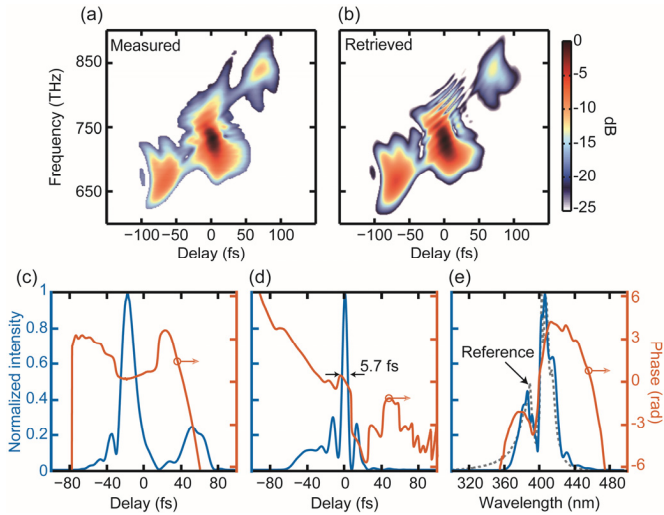

Fig. 3. (a) Measured SD-FROG trace of the compressed pulse. (b) Retrieved trace of (a). (c) Retrieved temporal profile and phase. (d) Numerically back-propagated temporal profile and phase from (c). (e) Corresponding spectral profile and phase of (d). The gray-dotted curve marks the reference spectrum of the pulse measured before the SD-FROG.

An interesting feature of self-compressing solitons is their potential to resonantly shed light to dispersive waves (DWs) due to the presence of higher-order dispersion. The frequency of such DWs is tunable over a wide spectral range from the VUV to the visible[18] by changing the gas species and pressure. In Fig. 4 we show the first examples of DW emission from UV solitons, tuned to 195, 220, and 260 nm (under-shaded with green, blue, and red, respectively). The influence of a core-wall resonance is clearly seen as a dip in the DW at ~220 nm (blue curve), in agreement with the 93 nm core-wall thickness (Fig. 2(d)).

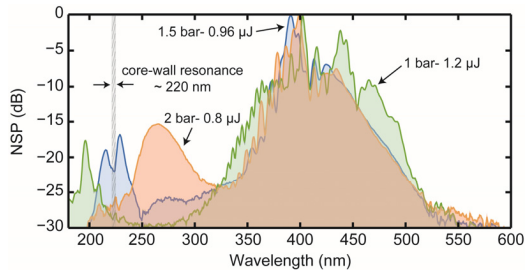

Fig. 4. Experimental results showing three DWs emitted at wavelengths from 195 to 260 nm in an Ar-filled HC-PCF. The vertical axis represents the spectral power normalized to its maximum (NSP). The position of the first core-wall resonance is indicated by the gray-shaded region.

## B. Ultraviolet supercontinuum generation

As mentioned in the introduction, broadband light sources extending from the DUV to the visible with high spectral power density would find many applications in areas such as photochemistry and spectroscopy. We have investigated the generation of a bright UV supercontinuum in a HC-PCF filled with either Ar or $H_2$. These gases have excellent transparency in the UV, while featuring very similar dispersion and Kerr nonlinearity, which makes comparisons between them easier and highlights the role of Raman scattering in $H_2$. Since pulses undergo extreme self-compression in HC-PCF, as shown in the preceding section, the impulsive response of the molecular gas can be exploited to further extend the supercontinuum.[19] The Raman response of $H_2$ is more pronounced in the UV owing to a dramatic enhancement in the Raman scattering cross-section; as a result the steady-state Raman gain, measured with narrowband pulses, is ~3 times higher at 400 nm than at 800 nm.[32]

Fig. 5 plots the supercontinua measured at the fiber output for a launched pump energy of 1.5 μJ and increasing pressure of Ar (Fig. 5(a)) and $H_2$ (Fig. 5(b)). Below 3 bar the supercontinuum extends over ~1.6 octaves for both gases. As the pressure increases, the zero-dispersion wavelength (ZDW) shifts to longer wavelength (white-dashed line), and the pump enters the normal dispersion regime at 4 bar for Ar and 5 bar for $H_2$. The absence of solitonic effects in the normal dispersion regime causes the pulse to broaden in time as it propagates, reducing its peak power and thus the efficiency of spectral broadening. In addition, self-steepening, which enhances the high-frequency content of the continuum, is much weaker in the normal dispersion region.[20] All this is clearly seen in the Ar system, where the spectrum significantly shrinks as the pressure increases (lower panel in Fig. 5(a)).

The situation is radically different when $H_2$ is used. Even though the electronic nonlinear refractive index of $H_2$ is ~1.6 times lower than that of Ar,[33] for pressures of ~2 bar the generated supercontinuum is broader. This is due to the Raman contribution to the quasi-instantaneous electronic polarization, which for $H_2$ includes rotational (S(1), period 57 fs) and vibrational (Q(1), period 8 fs) modes. For non-impulsive excitation in the transient regime, when the pulse duration is less than the coherence lifetime $T_2$ but greater than 57 fs, the Raman-related nonlinear refractive index is quasi-instantaneous and Kerr-like, resulting in SPM enhancement.[19] For pump pulse durations shorter than these oscillation periods, additional impulsive excitation of molecular motion can occur, causing a refractive index oscillation that can act as a phase-modulator for times less than $T_2$.[34] In the current work the pump pulses are always short enough to impulsively excite rotational motion, while for impulsive excitation of vibrational motion, self-compression below 8 fs is necessary. The resulting molecular motion creates Raman sidebands[35] that can themselves undergo SPM, producing additional spectral broadening. This mechanism is dominant for pressures above 5 bar in Fig. 5(b) (lower panel).

In the upper panels of Figs. 5(a)-(b) the measured output spectra for 10 bar of Ar and $H_2$ are plotted. The broader and flatter spectrum (8 dB limit from 280 to 700 nm) for $H_2$ illustrates the benefits of using a Raman-active gas. We also performed numerical simulations using the unidirectional field equation that describes the evolution of the forward-propagating electric field along the fiber.[36] The nonlinear polarization of the medium includes contributions from both Kerr and Raman effects, as well as photoionization. The

Raman polarization is modeled through a set of Maxwell-Bloch equations governing the excitation of molecular coherence and takes into account both rotational and vibrational Raman modes of hydrogen.[37] Other parameters such as polarizabilities, coherence lifetimes, and the nonlinear refractive indices of Ar and $H_2$ can be found in literature.[19,38,39] The retrieved pump pulse (shown in Fig. 2(c)) is used as initial condition for the simulations. To find the best agreement with the recorded spectra, the input pulse energy is decreased by ~30% compared to the experiment.

The green-shaded area in Fig. 5(b) shows the simulated spectrum for $H_2$ when the Raman contribution is switched off; the spectral width is comparable to that for Ar (Fig. 5(a)). Because of the strong rotational coherence that is excited impulsively, the high-frequency part of the supercontinuum which is mainly allocated in the trailing edge of the pulse, undergoes strong Raman-driven phase-modulation generating new sidebands. The interference between these sidebands can lead to structured output spectra as can be seen in the orange shaded area in Fig. 5(b). In addition, at this pressure a major part of the supercontinuum extends to the anomalous dispersion region, which together with the strong nonlinearities initiates ultrafast modulational instability. This combined with the Raman self-frequency shift can degrade the pulse-to-pulse coherence.[40] However, when averaged over many shots, these fine structures are smeared out resulting in a smoother continuum (the solid blue curve in Fig. 5(b)).

Moreover, photoionization does not significantly influence the simulated results, despite peak pulse intensities as high as $10^{14}$ W/cm$^2$.

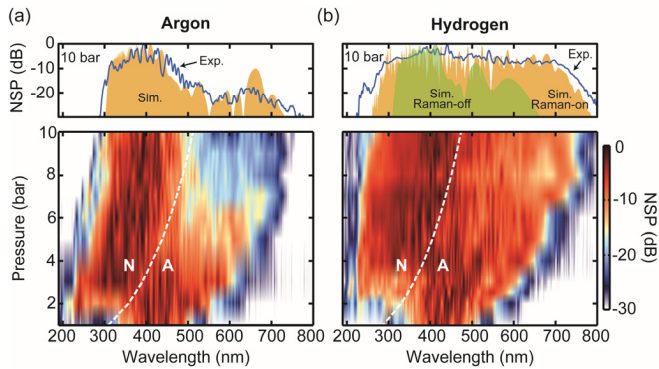

Fig. 5. UV supercontinua generated in (a) Ar and (b) $H_2$. Upper panels: experimental (blue lines) and simulated (shaded area) spectra generated by 1.5 μJ pulses for 10 bar of gas pressure. The green shaded area in (b) shows the simulated spectrum in the absence of Raman effects. Lower panels: spectra recorded for 1.5 μJ pump pulses with increasing gas pressure. The white-dashed line marks the ZDW. The recorded spectra are averaged over 400 consecutive pulses.

As the pressure is raised, hydrogen shows increasingly superior spectral broadening even for low pump energy, as shown in the lower panels of Fig. 6. Whereas at 1.3 bar and 720 nJ the spectra are comparable, at 9 bar the dispersion is normal and the advantages of using $H_2$ are apparent. The coherence of the generated spectra in the case of $H_2$ is numerically studied by calculating the modulus of the complex degree of first-order coherence ( $\left|g_{12}^{(1)}(\omega)\right| = \left|\left\langle E_1^*(\omega)E_2(\omega)\right\rangle \Big/ \sqrt{\left\langle\left|E_1(\omega)\right|^2\right\rangle\left\langle\left|E_2(\omega)\right|^2\right\rangle}\right|$, where the angle brackets denote an average over independent simulations) using an ensemble of simulations including pump pulses with different initial random noise.[15,40] The results shown by the solid green curves in the upper panels of Fig. 6 suggest that the entire spectrum remains coherent (i.e. $g_{12}^{(1)} \sim 1$). The simulated spectra are also shown with the light-brown curves.

A remarkable feature of the coherent spectra in Fig. 6(b) is that, despite a pump pulse energy of only 280 nJ, they have a bandwidth sufficient to support ~4.6 fs pulses for Ar and ~3.8 fs pulses for $H_2$. The pump energy can be further reduced by increasing the gas pressure or reducing the core diameter. This opens a new route to producing frequency combs in the UV-visible region using frequency-doubled femtosecond laser pulses from commercial near-infrared mode-locked oscillators operating at tens of MHz repetition rates.

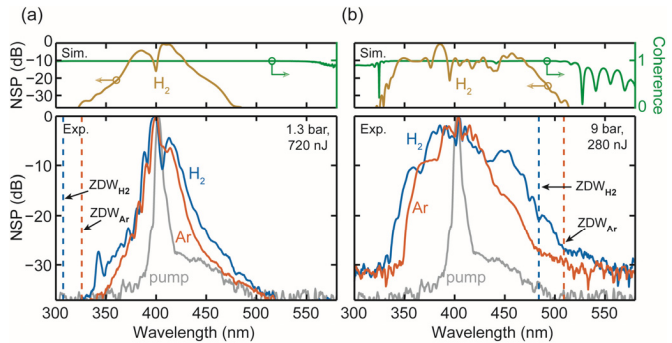

Fig. 6. Spectra for two different gas pressures and pulse energies of (a) 1.3 bar and 720 nJ and (b) 9 bar and 280 nJ. Upper panels: simulated spectrum of $H_2$ (light-brown curve) and the corresponding first-order coherence (green curve). Lower panels: recorded spectra using $H_2$ (blue curves) and Ar (orange curves). The vertical-dashed lines mark the positions of the ZDWs. The gray curves show the spectra transmitted in the evacuated fiber. The parameters in (a) yield soliton orders of 9 for Ar and 6.5 for $H_2$. In (b) the pump lies in the normal dispersion region."

## CONCLUSIONS

Gas-filled HC-PCF is highly suited to efficient ultrafast nonlinear optics in the UV. For example, pulses of duration ~40 fs and energy 520 nJ at 400 nm can be compressed to sub-6-fs using soliton dynamics in an Ar-filled fiber. These self-compressed UV solitons emit tunable dispersive waves in the VUV-DUV when launched in small-core HC-PCFs with sufficiently thin core-walls. Using hydrogen as filling gas, flat DUV-to-visible supercontinua can be generated, yielding much broader and flatter spectra than in an argon-filled fiber, owing to the enhanced Raman response of $H_2$ in the UV. Ultrafast UV sources based on gas-filled HC-PCF are likely to be of great interest in spectroscopy, photochemistry and studies of molecular dynamics.